\documentstyle[prl,aps,twocolumn,epsf]{revtex}
\newcommand{\raf}[1]{(\ref{#1})}
\def\lsim{ \,\, \vcenter{\hbox{$\buildrel{\displaystyle <}\over\sim$}}
 \,\,}
\def\gsim{ \,\, \vcenter{\hbox{$\buildrel{\displaystyle >}\over\sim$}}
 \,\,}
\newcommand{\be}{\begin{eqnarray}}
\newcommand{\ee}{\end{eqnarray}}
\newcommand{\non}{\nonumber\\}

\newcommand{\ave}[1]{\left\langle #1 \right\rangle}
\newcommand{\absol}[1]{\left| #1 \right|}
\begin{document}
\twocolumn[\hsize\textwidth\columnwidth\hsize\csname
@twocolumnfalse\endcsname
\hfill LBL-47721 

\hfill Mc-Gill-preprint-01-09 

\hfill GSI-Preprint/2001-10

\ \\

\title{
Fluctuations of rare particles as a measure of chemical equilibration
}
\author{
S. Jeon$^{a}{}^\dagger{}^*$,
V. Koch$^a$, K. Redlich$^{a,c}$, and X.-N. Wang$^a$
}
\address{
        $^a$Nuclear Science Division,
Lawrence Berkeley National Laboratory, 1 Cyclotron Road,
Berkeley, CA 94720}
\address{
$^c$ GSI, D-64291 Darmstadt,
Germany and Institute of Theoretical Physics, University of
Wroc\l aw, PL-50204 Wroc\l aw, Poland} 
\address{$\dagger$Department of Physics, McGill
University, 3600 University Street, Montreal, QC H3A-2T8, Canada}
\address{$*$RIKEN-BNL Research Center, Brookhaven National Laboratory,
Upton, NY 11973}
\date{\today}
\maketitle

\begin{abstract}
We calculate the time evolution of fluctuations for rare particles such as e.g.
kaons in 1 AGeV  or charmonium in 200 AGeV heavy ion collisions.
We find that these fluctuations are a very sensitive probe of the
degree of chemical equilibration reached in these collisions.
Furthermore, measuring the second factorial moment  the size of
the initial population can be determined. \vspace{0.5in}
\end{abstract}
]

\section{Introduction}

Statistical models have long been used as a tool to describe
particle production in heavy ion and in high energy particle
collisions \cite{Hag71,sur,he}. Recent analysis have shown that
these models can indeed give a satisfactory description of the
multiplicities of most hadrons measured in A-A collisions for
bombarding energies ranging from 1 AGeV (SIS) to 160 AGeV (SPS)
\cite{Bra99,our}. Especially at the low energies (SIS) and for the
rare particles such as kaons, the success of the statistical
description has been a puzzle and could not be understood within
the state-of-the-art transport models \cite{brat}. For instance a
recent analysis for chemical equilibration within a transport
model \cite{brat} gives chemical equilibration time of the order
of 300 fm/c for a situation relevant to 1 AGeV heavy ion
collisions, considerably larger than the typical duration of such
reaction, which is of the order of 30-50 fm/c. On the other hand
the success of the statistical model cannot be disputed. Not only
are the particle ratios reproduced for central collision over a
wide range of bombarding energy, but also the centrality
dependence at low energies is consistent with the predictions of
the statistical models. In particular the almost {\em quadratic}
dependence of the kaon multiplicity on the number of participating
nucleon at SIS follows directly from the statistical model, once
strangeness conservation is taken into account exactly
\cite{redlich}. Transport models on the other hand fail to
reproduce the observed centrality dependence and particularly energy
    dependence of the $K/\pi$-ratio \cite{cassing}.

In a recent paper \cite{kkl} some of us have shown, that the
chemical equilibration time is considerably shortened if the
strangeness conservation is taken into account explicitly.
However, using the new
rate equations derived  in this work combined with the cross
sections as given in \cite{tsushima} one still arrives at
equilibration time substantially exceeding the lifetime of the
system.

Does this mean that the success of the statistical model is a pure
coincidence? This is very unlikely  as statistical model
naturally explains most of the basic features of experimental data
in a very broad energy range from SIS up to SPS.
It is conceivable that there are additional processes at work,
like e.g. many particle collisions \cite{greiner} or in medium
modifications of hadron properties \cite{brown}, which are not yet
taken  into account in transport models.

Thus, a direct experimental determination of the rate of
equilibration in heavy ion collisions is called for as it 
would possibly provide evidence for new physics. In this paper we
will demonstrate that the fluctuations of rare particles is a very
sensitive probe of the degree of equilibration reached in these
collisions. Such a measurement, though certainly difficult, could
for the first time provide a direct experimental evidence for
chemical equilibration in heavy ion reactions.

This paper is organized as follows. In the following section we set up the
formalism. Then we present the results for the time dependence of the second
factorial moment for several initial conditions. Before we turn to
observational issues we also will discuss the case when no constraints from
any $U(1)$ charge conservation are present.

\section{Formalism}
In Ref. \cite{kkl} the rate equation for particles which are subject
to an explicit $U(1)$ ``charge'' conservation has been derived.
Considering a binary process $a_1 a_2 \rightarrow b_1 b_2$
with $a \neq b$ one arrives at the following master equation for the
probabilities $P_n$ to find $n$ particles 
\be
\frac{dP_n}{d \tau}&=&\epsilon \left [P_{n-1}-P_n \right] \nonumber
\\
&-&  \left [ n^2 P_n - (n+1)^2 P_{n+1} \right ], \label{generaln}
\ee 
where $n=0,1,2,3,\cdots$. Here 
\be \epsilon \equiv G \langle
N_{a_1} \rangle \langle N_{a_2}\rangle /L, 
\ee 
and the
dimensionless time variable $\tau$ is defined as 
\be \tau = t
\frac{L}{V} 
\ee 
so that $\tau$ is measured in units of the
relaxation time $\tau_0^C = V/L$ \cite{kkl}. The momentum-averaged
cross sections for the gain process $a_1 a_2 \rightarrow b_1 b_2$
and the loss process $b_1 b_2 \rightarrow a_1 a_2$ are defined as
$G \equiv \langle \sigma_G v \rangle$ and $L \equiv \langle
\sigma_L v \rangle $, respectively. The ratio of these momentum
averaged cross sections is related to the ratio of equilibrium particle
densities involved 
\be 
\frac{G}{L}= \frac {d_{b_1} \alpha_{b_1}^2
K_2 (\alpha_{b_1}) d_{b_2} \alpha_{b_2}^2 K_2 (\alpha_{b_2})}
{d_{a_1} \alpha_{a_1}^2 K_2 (\alpha_{a_1}) d_{a_2} \alpha_{a_2}^2
K_2 (\alpha_{a_2})}, 
\ee
where  $d_k$'s denote the degeneracy
factors, and $\alpha_k \equiv m_k/T$.

Eq. \raf{generaln} has no obvious solution and needs to be
solved numerically. The asymptotic (equilibrium)
probability distribution, on the other hand, has
been derived in \cite{kkl}
\be
P_{n,\rm eq.}=\frac{\epsilon^n}{I_0 ( 2\sqrt \epsilon )~(n!)^2}.
\label{eq:p_equil}
\ee
leading to
\be
\ave{N}_{\rm eq.} &=& \sqrt{\epsilon}
\frac{I_1 ( 2\sqrt \epsilon )}{I_0 ( 2\sqrt \epsilon )} =
\epsilon
- {\epsilon^2\over 2}
+ {\epsilon^3\over 3} + \ldots
\label{n_eq}
\\
\ave{N^2}_{\rm eq.} &=& \epsilon.
\label{n2_eq}
\ee

The above general rate equation is valid for arbitrary values of
$\langle N \rangle$ for particle production constrained by $U(1)$
charge conservation. It reduces to the grand canonical results for
large $\langle N \rangle$ and to the canonical results for small
$\langle N \rangle$. It provides a generalization of the standard
rate equation beyond the grand canonical limit. It was shown
\cite{kkl} that for rare particle production the equilibrium
abundance is  much smaller and the relaxation time is much shorter
than expected from the standard rate equation. In this paper we
will discuss further consequences of the 
generalized  rate equation  and in
particular study the time evolution of the multiplicity
fluctuations. We want to demonstrate that the combined information
on both $\langle N\rangle$ and 
$\langle N^2\rangle$ can help to determine the degree
of chemical equilibration.

\section{Results}
\label{sec:with_u1}

In this work we will be mostly concerned with the fluctuations of the
particle number
in the case of rare particle, i.e. $\ave{N}_{\rm eq.} \ll 1$ or,
equivalently, $\epsilon
\ll 1$. In particular we will investigate the behavior of the second
factorial moment $F_2$
\be
F_2 \equiv \frac{\ave{N(N-1)}}{\ave{N}^2}.
\ee
{}From Eqs. \raf{n_eq} and \raf{n2_eq} in the limit of small $\epsilon$ the
equilibrium value for $F_2$ is given by
\be
F_2 = \frac{1}{2} + \frac{\epsilon}{6} + \ldots =
\frac{1}{2} + \frac{\ave{N}_{\rm eq.}}{6} + \ldots
\ee

In order for the second factorial moment to be a sensitive probe of the degree
of equilibrium achieved, one needs to investigate its initial value. Here we
consider two distinct cases:
\begin{enumerate}
\item The initial particle number is considerably smaller than the equilibrium
  value. This is relevant, for example, for kaon production in 1~AGeV heavy ion
  collisions.
\item
  The initial particle number is considerably larger than the equilibrium
  value. This might be relevant for charm production in 200~AGeV heavy
  ion collisions \cite{Stachel_charm}
\end{enumerate}
In the first case, where the initial particle number is small, let us consider
two scenarios. On one hand, let us assume that initially the probabilities
$P_n$ are distributed according to a Poisson distribution:
\be
P_n(\tau=0) = \frac{N_0^n}{n!}e^{-N_0},
\ee
where $N_0$ is the initial average number of particles.
In this case, the factorial moment obviously starts out at
\be
F_2(\tau=0) = 1
\ee
and decreases by a factor of two until equilibrium is reached.
On the other hand one may assume that initially there is at most one particle
in a given event. In this case the initial conditions are
\be
P_0(\tau=0) &=& 1-N_0,\nonumber \\
P_1(\tau=0) &=& N_0, \nonumber \\
P_n(\tau=0) &=& 0 \,\,\,\, n>1,
\label{binomial}
\ee
which we will refer to as `binomial' initial conditions.
As shown in Appendix~\ref{app:init_time},
$F_2$ starts out at $F_2 = 0$, but almost immediately
reaches a maximum after a time of the order of
\be
\tau_{\rm max} \simeq \frac{N_0}{N_{\rm eq.}}
\ee
and for $N_0/N_{\rm eq.} \ll 1$, $F_2^{\rm max} \simeq 1$.

\begin{figure}[htb]
  \setlength{\unitlength}{1cm}
  \epsfxsize=8cm
  \centerline{\epsfbox{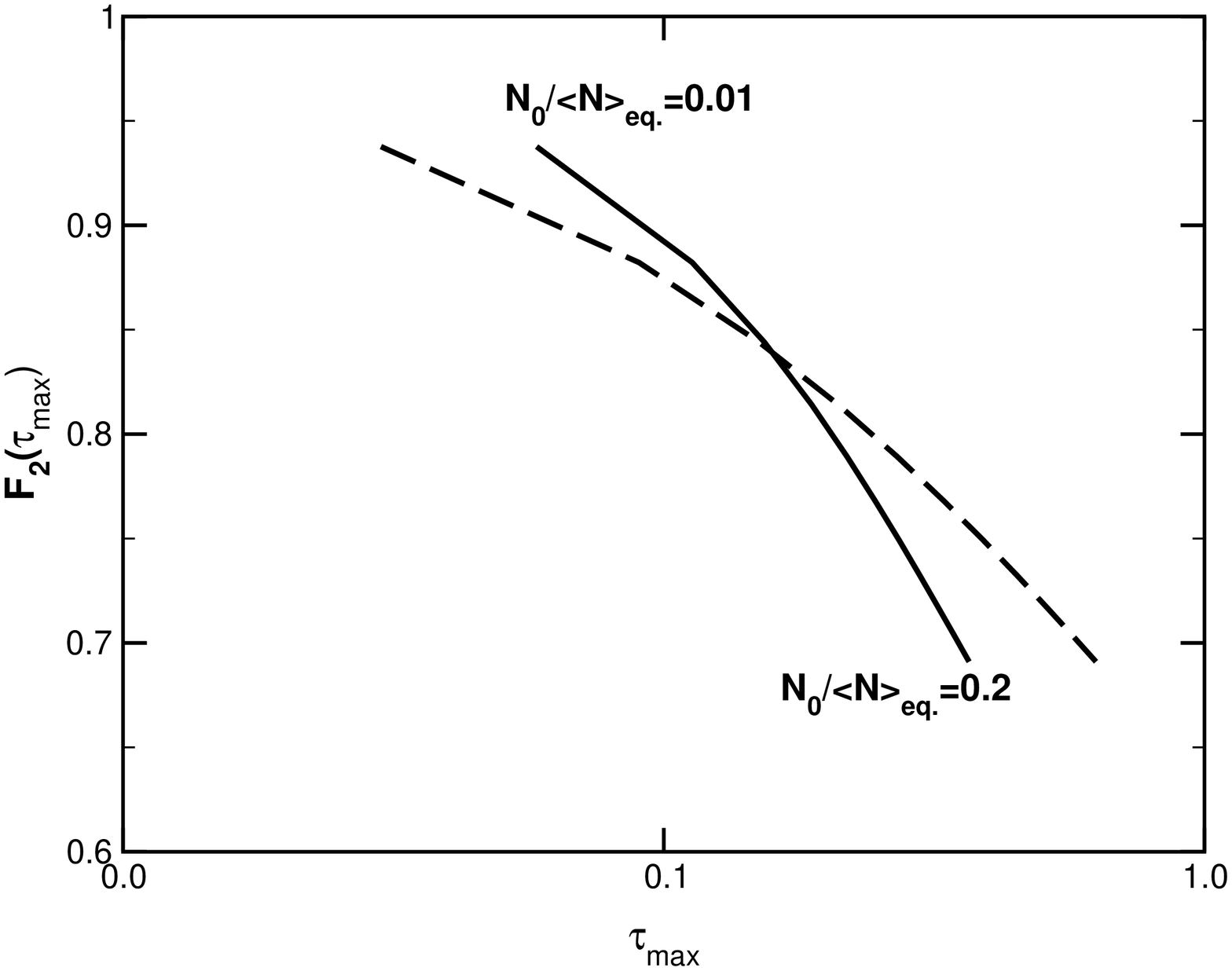}}
  \caption{Location and value of the maximum of $F_2(\tau)$ for a range of
    values of $N_0/N_{\rm eq.}$ as indicated in the figure.
    Here $\epsilon = 0.1$
    has been used. The dashed line assumes that
    $\tau_{\rm max} = 3 N_0/N_{\rm eq.}$ }
  \label{fig:maximum}
\end{figure}

Therefore $F_2$ approaches equilibrium from above and a
measurement of $F_2 > 1/2$ will indicate the degree of equilibrium
that has been reached in a heavy ion collision.
The detailed time and height, where $F_2$ reaches a maximum depend, of course,
on the input parameters $N_0$ and $\epsilon$. The dependence of
$F_2^{\rm max}$ on the ratio
$N_0/N_{\rm eq.}$ for $\epsilon = 0.1$ is shown in Fig.~\ref{fig:maximum} 
as the full line. The
dashed line in Fig.~\ref{fig:maximum} is obtained by assuming
$\tau_{\rm max} = 3 N_0 / N_{\rm eq.}$ showing that indeed the time scale for
reaching the maximum is given by $N_0 / N_{\rm eq.}$.

We further see that for small $N_0$ the factorial moment
essentially immediately reaches a value close to $F_2 = 1$, giving
a factor of two sensitivity on the degree of non-equilibrium
established  in the collisions. Obviously, in the case where $N_0
\simeq \epsilon = N_{\rm eq.}$ this sensitivity is lost, as the
equilibrium value is very close to the initial value.

\begin{figure}[htb]
  \setlength{\unitlength}{1cm}
  \epsfxsize=8cm
  \centerline{\epsfbox{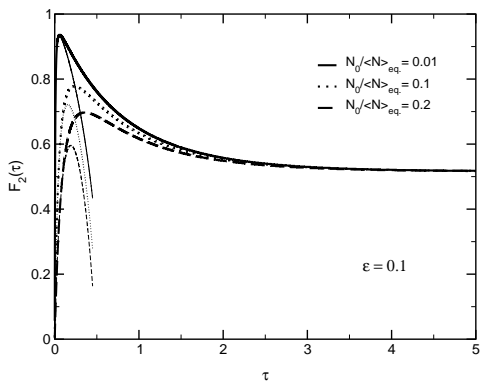}}
  \caption{Time evolution of the factorial moment $F_2$ for several initial
    particle numbers $N_0$ (thick lines). The thin lines show the result of the
    approximate formula \raf{approx_1}. Here $\epsilon=0.1$ has been used.}
  \label{fig:evolve_1}
\end{figure}

Assuming  binomial initial conditions in Fig.~\ref{fig:evolve_1}
we show the full time evolution for several initial particle numbers.
For small times the
approximate solution \raf{approx_1} is also shown. Clearly, the equilibrium
value of $F_2$ is reached from above, but the effect becomes small as the
initial particle number becomes comparable with the equilibrium value.

\begin{figure}[htb]
  \setlength{\unitlength}{1cm}
  \epsfxsize=8cm
  \centerline{\epsfbox{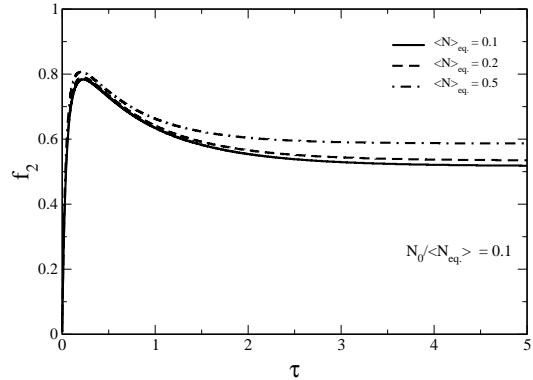}}
  \caption{Time evolution of the factorial moment $F_2$ for several
  equilibrium numbers or equivalently values of $\epsilon$. Here binomial
    initial conditions with $N_0/N_{\rm eq.} = 0.1$ have been chosen.}
  \label{fig:evolve_eps_compare}
\end{figure}

In Fig.~\ref{fig:evolve_eps_compare} we show the time evolution of $F_2$ 
for different choices of $\epsilon$ or, equivalently, for different
equilibrium particle numbers. Obviously, the larger the equilibrium particle
number is the smaller is the effect.

For the second case when the initial population is much bigger
than the equilibrium population,
then the annihilation process must dominate during the early
stage of evolution for a short period of time.
In this case, one can look for a perturbative solution around
$\epsilon = 0$.
Since there can be many different initial conditions with a large
initial population, it is better here to use the generating function\cite{kkl}
\be
g(\tau, x) = \sum_{n=0}^\infty x^n P_n(\tau),
\ee
and the equation it satisfies
\be
{\partial g \over \partial \tau} =
(1-x)\left( x g'' + g' - \epsilon g \right)
\;.
\label{eq:gen_func}
\ee
where the prime indicates a derivative with respect to $x$.
The averages needed to calculate the first two factorial moments
are given by
 \be
 \ave{N} = g'(\tau, 1)\ \ \ \hbox{and}\ \ \
 \ave{N(N-1)} = g''(\tau, 1).
 \label{eq:aves}
 \ee
 Details of the perturbative procedure can be found
 in Appendix~\ref{app:pert}.
 To understand the qualitative picture, first consider the initial time
 $\tau=0$.  Since we assume $\ave{N} \gg 1$ initially,
 we must have $\ave{N(N{-1})} = {\cal O}(\ave{N}^2)$ and hence,
 $F_2 = {\cal O}(1)$ at $\tau = 0$.
 For $\tau \gsim 1$, the first order perturbative solutions are
\be
& &
\ave{N}
=
\ave{N}_{\rm eq.} + \absol{a_1} e^{-\tau}
+ {\cal O}(e^{-4\tau}),
\label{eq:pert_ave}
\\
& &
\ave{N(N{-}1)} =
\ave{N(N{-}1)}_{\rm eq.} + {{{\frac{2\,\epsilon}{5}}  \absol{a_1}}{{e^{-\tau}}}}
+ {\cal O}(e^{-4\tau}),
\ee
 where $a_1$ is a ${\cal O}(1)$
 constant determined by the initial condition.  From 
 the above equations and  using Eqs.~(\ref{n_eq}) and
(\ref{n2_eq}) the second factorial moment is then given by
 \be
 F_2(\tau)
 \simeq
 {1\over 2}{
 \epsilon^2 + (4/5)\epsilon \absol{a_1}e^{-\tau}
 \over
 \left(\epsilon + \absol{a_1} e^{-\tau}\right)^2
 }.
\label{eq:pert_f2}
 \ee
 For times $1 \lsim \tau \lsim {-}\ln\epsilon$,
 the exponential terms in Eq.~(\ref{eq:pert_f2}) dominate.
 As a result, $F_2(\tau) = {\cal O}(\epsilon)$
 within the interval $1 \lsim \tau \lsim {-}\ln\epsilon$.
 Note that for arbitrarily small $\epsilon$, this interval can be arbitrarily
 long.  Also, since
 ${\cal O}(\epsilon) \ll F_2(0)$ and ${\cal O}(\epsilon) \ll F_2^{\rm eq.}$,
 $F_2$ must reach a minimum somewhere inside that interval.

\begin{figure}[htb]
 \setlength{\unitlength}{1cm}
  \epsfxsize=8cm
  \centerline{\epsfbox{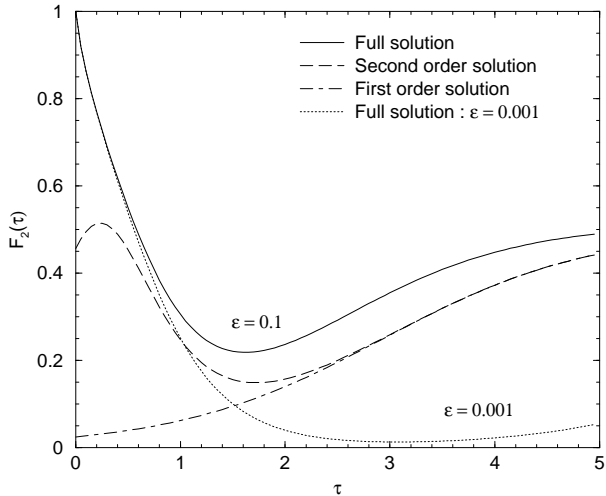}}
  \caption{The second factorial moment as a function of
  time.  Initial distribution is a Poisson distribution with 
  $N_0 = 5$.  $\epsilon = 0.1$.
  The solid line represents the numerical solution.  The dashed line is the
  result of the second order perturbative calculation and the dot-dashed line
  is the result of the first order perturbative calculation.  Also shown
  is the full numerical solution with the same initial condition
  and $\epsilon = 0.001$.
}
  \label{fig:pert_sol}
 \end{figure}

For illustration, let us choose a Poissonian initial distribution
 with $N_0 = 5$ and    also set $\epsilon = 0.1$.
 The numerical solutions and
 Eq.~(\ref{eq:pert_f2}) as well as the second order perturbative solution
 are displayed in Fig.~\ref{fig:pert_sol}.
 The numerical solution clearly shows the rapid initial decrease and the
 subsequent slower rise to the equilibrium value.
 To illustrate the duration of the small $F_2$ interval, we also show
 the full numerical result with $\epsilon = 0.001$.  The longer period of time
 where $F_2$ stays to be ${\cal O}(\epsilon)$ is clearly visible.

 To sum up, $F_2$ as a function of time must:
 \begin{enumerate}
 \item[(i)] Start from ${\cal O}(1)$.
 \item[(ii)] Reach the minimum value of ${\cal O}(\epsilon)$ 
at $\tau \sim 1$ which is
 much smaller than $F_2^{\rm eq.} = 1/2 + {\cal O}(\epsilon)$.
 \item[(iii)] Stay ${\cal O}(\epsilon)$ until
 $\tau \sim -\ln\epsilon$.
 \item[(iv)] Approach the equilibrium value from below after
 $\tau \sim -\ln\epsilon$.
 This is in contrast to the first case where we considered
 $N_0 \ll \ave{N}_{\rm eq.}$.
 \end{enumerate}
 Hence, if experimental value of $F_2$ is smaller than $1/2$, then it is
 a strong indication that the equilibrium is not reached and furthermore
 it also indicates that
 the initial population was much larger than the equilibrium one.


\section{Absence of $U(1)$-charge conservation}
\label{sec:no_u1}

It is  interesting to also study particle production without
the constraint of $U(1)$ charge conservation.  Obviously the time
evolution equation for the multiplicity distribution should be
different from what we have described so far  for strongly
correlated processes. Let us consider a general process $a+b
\leftrightarrow c+d$ without any constraint of charge
conservation. We denote $P_n$ as the multiplicity distribution for
particle $c$. It then satisfies the following evolution equation,
\begin{equation}
  \frac{dP_n}{d\tilde{\tau}}=\tilde{\epsilon} [ P_{n-1}-P_n] -[nP_n-(n+1)P_{n+1}],
  \label{generaln_no_u1}
\end{equation}
where $\tilde{\tau}= \tau / \ave{N_d} = tL/V/\langle N_d\rangle$
is a scaled time and ${\tilde\epsilon}  = \epsilon / \ave{N_d} =
G\langle N_a\rangle \langle N_b\rangle/L/\langle N_d\rangle$.

The generating function $g(\tilde{\tau}, x)$ for $P_n$
\be
g(\tilde{\tau}, x) = \sum_{n=0}^{\infty} x^n P_n(\tilde{\tau})
\ee
satisfies the following partial
differential equation,
\begin{equation}
  \frac{\partial g}{\partial \tilde{\tau}}=(1-x)[g^\prime -\tilde{\epsilon} g].
\label{grand}
\end{equation}
It is interesting to note that the above equation is very similar
to Eq.~(\ref{eq:gen_func}) for a constrained system except that
it does not contain the second derivative on the right-hand side.
Therefore, for certain period during which one can neglect the
second derivative of the generating functions, the evolution of the
multiplicity distribution in a canonical system should be similar
to that of a grand canonical.

The general solution to the Eq.~(\ref{grand}) can be found
\begin{equation}
  g(\tilde{\tau}, x)=g_0((1-x)e^{-\tilde{\tau}})e^{\tilde{\epsilon}(1-x)(e^{-\tilde{\tau}}-1)}
  \label{solu}
\end{equation}
with the initial condition $g(0, x)\equiv g_0(1-x)$.
The normalization condition
$g(\tilde{\tau}, x=1)=\sum P_n=1$ also implies $g_0(0)=1$. One can readily find the
equilibrium solution in the limit $\tilde{\tau}=\infty$,
\begin{equation}
 g_{\rm eq.}(x)=e^{-\tilde{\epsilon}(1-x)} \label{eq}
\end{equation}
with the corresponding equilibrium multiplicity distribution

\begin{equation}
  P_{n,{\rm eq.}}=\frac{\tilde{\epsilon}^n}{n!}e^{-\tilde{\epsilon}},
\end{equation}
which is a Poisson distribution with averaged multiplicity
$\langle N\rangle_{\rm eq.}=\tilde{\epsilon}$. One can also easily
calculate the first and second factorial moments of the multiplicity
distribution,
\begin{eqnarray}
\langle N\rangle &=&\epsilon +(\langle N\rangle_0-\epsilon)e^{-\tilde{\tau}} \\
\langle N(N-1)\rangle&=&\langle N(N-1)\rangle_0 e^{-2\tilde{\tau}}\nonumber\\
&-&2\epsilon \langle N\rangle_0 e^{-\tilde{\tau}}(e^{-\tilde{\tau}}-1)
+\epsilon^2(e^{-\tilde{\tau}}-1)^2.
\end{eqnarray}
They both approach to their equilibrium values exponentially.

One interesting case of a special
initial condition is $g_0=1$ when the initial multiplicity
$\langle N\rangle_0$ is zero. Comparing
Eqs.({\ref{solu}) and (\ref{eq}), one finds that the multiplicity
distribution in this case remains a Poissonian, and consequently the
factorial moment $F_2 = 1$ at all times.
For the binomial initial conditions (Eq.\raf{binomial}),
\be
g(\tilde{\tau}=0, x) = 1 - N_0 (1-x),
\ee
the factorial moment $F_2$ starts out from $F_2(\tilde{\tau}=0)=0$, but
approaches the equilibrium value via
\be
F_2 = 1-[N_0/( \tilde{\epsilon}(e^{\tilde{\tau}}-1) + N_0)]^2
\ee
at a time
scale of
\be
\tilde{\tau} \simeq \ln (1+\sqrt{2}N_0/\tilde{\epsilon}) \simeq
\sqrt{2}N_0/\tilde{\epsilon}.
\ee

Comparing the results presented here with the previous  section it
is clear that an additional constraints imposed by the U(1) charge
conservation laws are implying a  crucial modification  of not
only the equilibrium probability distributions of particle number
but also their fluctuations and  time evolution towards the
equilibrium limit.

\section{Towards experimental observables}
So far we have discussed the production of a single species of particles with
conserved quantum numbers, such as e.g. $K^+$-$K^-$-pairs. In reality,
however, one has to deal with more than one species. For example at 1~AGeV
heavy-ion collisions the relevant strange degrees of freedom are $K^+$ and
$K^0$ which carry positive strangeness and the $\Lambda$ and $\Sigma$ hyperons
carrying negative strangeness. The anti-kaons are not relevant in this case as
the ratio of $K^-/K^+$ is very small ($\simeq 2 \%$ at 1.5 AGeV
\cite{Oeschler} ). The results shown in the previous section thus apply to the
combined multiplicities for $K^+$ and $K^0$, i.e.  
\be 
N = N_{K^+} + N_{K^0}.
\ee
Very often, experiments can either measure $K^+$ or $K^0$ but not both species
at the same time. The above master equation has been extended for more than
one particle species carrying the conserved quantum number in reference
\cite{lin_ko}. The equation for the combined probability $P_{i,j}$ to find $i$
$K^+$ and $j$ $K^0$ mesons is given by 
\be \frac{d P_{i,j}}{d \tau} =
\epsilon_1 (P_{i-1,j} - P_{i,j}) + \epsilon_2 \frac{L_2}{L_1} (P_{i,j-1} -
P_{i,j}) \non - (i(i+j) P_{i,j} - (i+1)(i+j+1)P_{i+1,j}) \non -
\frac{L_2}{L_1}(j(i+j) P_{i,j} - (j+1)(i+j+1) P_{i,j+1}),
\label{new_master1}
\ee 
where $\tau = t \frac{L_1}{V}$ and $\epsilon_{1,2} \equiv G_{1,2} \langle
N_{a_1} \rangle \langle N_{a_2}\rangle / L_{1,2}$. 
The equilibrium solution is given by \cite{lin_ko}
\be
P_{i,j}^{\rm eq.} = \frac{\epsilon_{tot}^{i+j}}{I_0(2 \sqrt{\epsilon_{tot}}) 
((i+j)!)^2}  
\frac{(i+j)! \epsilon_1^i \epsilon_2^j }{\epsilon_{tot}^{i+j} i! j!}
\label{new_equil}
\ee
with $\epsilon_{tot} \equiv \epsilon_1+\epsilon_2$. Note that the equilibrium
probability distribution is the product of the distribution of pairs according
to Eq. \raf{eq:p_equil} 
and a binomial distribution, which determines the relative
weight of the individual particles, in our case the $K^+$ and $K^0$.
For the equilibrium configuration the relevant expectation values are then 
easily computed
\be
\ave{N_1}_{\rm eq.} &=& f_1  \ave{N}_{\epsilon_{tot}}, 
\\
\ave{N_1^2}_{\rm eq.} &=& f_1^2 \ave{N^2}_{\epsilon_{tot}} + 
f_1(1-f_1) \ave{N}_{\epsilon_{tot}} ,
\\
F_2^{equil}(K^+) &=& \frac{1}{2} + \frac{\epsilon_{tot}}{6} + \ldots
\ee
where $f_1 = \epsilon_1/\epsilon_{tot}$. The average $\ave{}_{\epsilon_{tot}}$
denotes the
averages given in Eqs. \raf{n_eq} and \raf{n2_eq} 
with $\epsilon \rightarrow \epsilon_{tot}$ For small $\epsilon_{1,2}$ 
the effect of the second species only appears at next to leading order in
$F_2$.  
      
The master equation governing the particle of interest, say the $K^+$ is
obtained by summing equation \raf{new_master1} over the index of the particle
which is not observed.
\be 
\frac{d P_{i}}{d \tau} &=&
\epsilon_1 (P_{i-1j} - P_{i}) \non 
&&- (i^2 P_{i} - (i+1)^2 P_{i+1}) \non 
&& - (i \sum_j j P_{i,j} - (i+1) \sum_j j P_{i+1,j})
\label{new_master2}
\ee 
with 
\be
P_{i} \equiv \sum_j P_{i,j}.
\ee

Comparing with the original equation \raf{generaln}, the presence of the 
other species, the $K^0$ in our case, leads to the last two terms of equation
\raf{new_master2}. However, in the situation of interest here, where $N_{K^0}
\ll 1$ these terms can be neglected. Thus we recover the original equation
governing the time evolution of the $K^+$. This can also be seen in
Fig. \raf{fig:2_flavor} where we compare the evolution of $F_2$ based on
Eq. \raf{generaln} with that based on Eq. \raf{new_master2}. 
For the case at hand, namely
kaon production in heavy ion collisions, isospin symmetry suggests that the
production and absorption cross-section for $K^+$ and $K^0$ are roughly the
same, i.e. $L_1 = L_2$ and $\epsilon_1 = \epsilon_2$.
Here we have assumed that due to isospin symmetry the 
collision rates for both kaon species are identical.

To summarize the effect of the second species leads to a sub-leading 
correction $\sim 
\epsilon_{K^0}/6 \simeq \ave{K^0}/6$ which in the present context is
negligible.

\begin{figure}[htb]
  \setlength{\unitlength}{1cm}
  \epsfxsize=8cm
  \centerline{\epsfbox{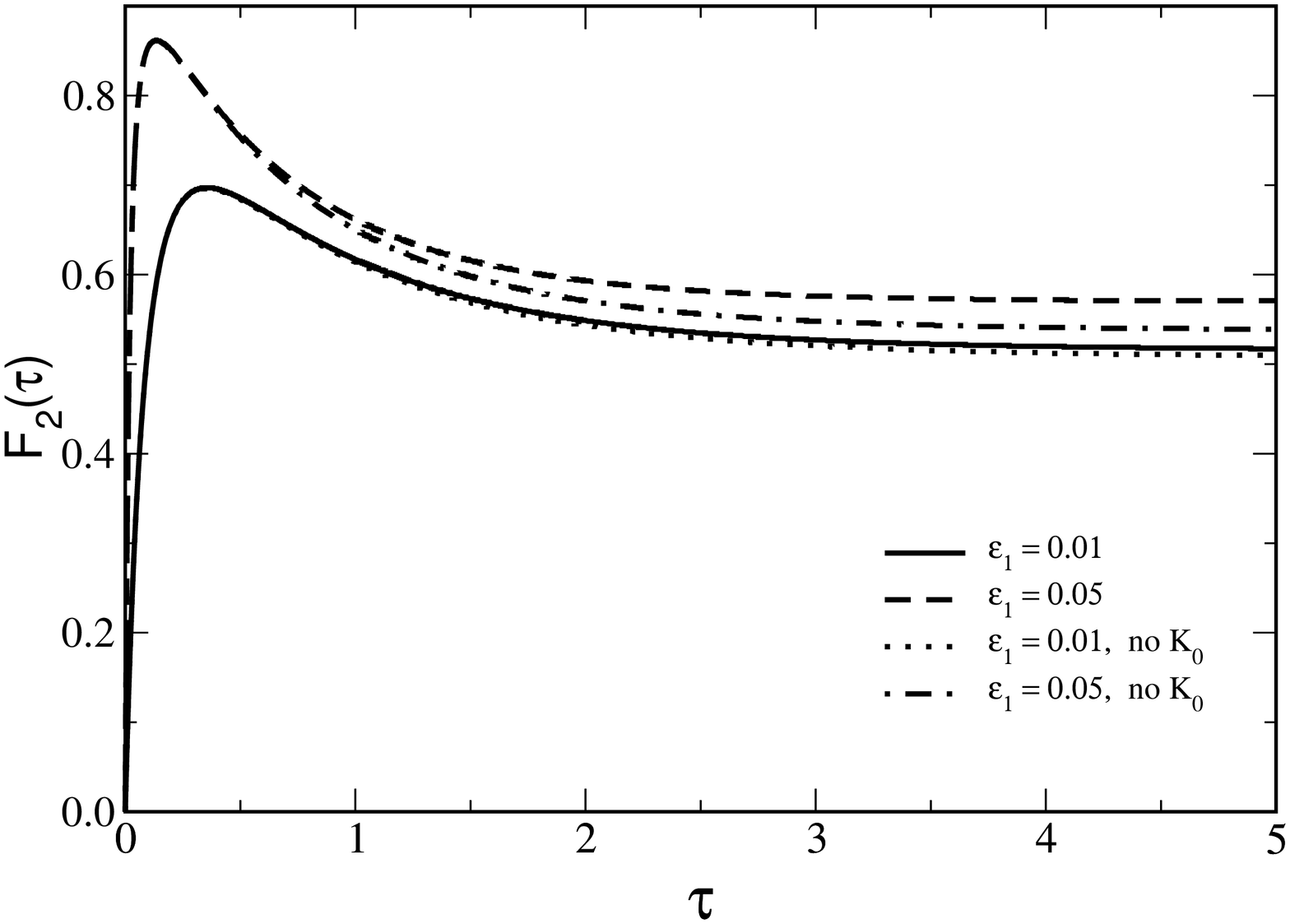}}
  \caption{Comparison of time evolution of $f_2$ for one and two particle
  species assuming binomial initial conditions and $N_0 = 0.01$}
  \label{fig:2_flavor}
\end{figure}

Let us now turn to the question of how to measure these
fluctuations in experiment. Since we are considering   rare
particles, the measurement of fluctuations are obviously 
difficult. Here we propose to study the second factorial
moment. Therefore, besides the inclusive particle number $\ave{N}$ one
also has to measure $ \ave{N(N-1)}$. The latter expectation
value, however, is directly related to the two particle density
\be 
\rho_2(p_1,p_2) = \frac{d^2N}{d^3 p_1\,d^3 p_2} 
\ee 

\be
\ave{N(N-1)} = \int d^3 p_1 \, d^3 p_2 \, \rho_2(p_1,p_2). 
\ee 

It is interesting to note (see also \cite{tannenbaum} )
that the same information enters the measurement of
so called HBT or Bose-Einstein (BE) correlations \cite{hbt_article}. 
The BE-correlation function as a function of the relative momentum is
defined as
\be
C_{BE}(q) = \frac{\rho_2(q)}{\rho_{11}(q)},
\ee
where 
\be
\rho_2(q) \equiv \int dp_1 dp_2 \rho_2(p_1,p_2) \delta(|p_1 - p_2|-q)
\ee
and
\be
\rho_{11}(q) \equiv \int d p_1 d p_2 \rho_1(p_1) \rho_2(p_2) 
\delta(|p_1 - p_2|-q)
\ee
Usually the $C_{BE}$ is parameterized as
\be
C_{BE}(q) = 1 + \lambda e^{-q^2 R^2}
\ee
so that outside the correlation region $q \gg 1/R_{source}$ the correlation
function assumes a value of one. However, in case of rare particle subject to 
$U(1)$ conservation law, i.e. kaons at low energy heavy-ion collisions, 
this will be different:

In terms of $\rho_2(q)$ and $\rho_{11}(q)$ the factorial moment $F_2$ 
is given by
\be
F_2 = \frac{\int dq \rho_2(q)}{\int dq \rho_{11}(q)} \rightarrow \frac{1}{2},
\label{f2_correct}
\ee
which, as shown,  assumes a value of $F_2 = 1/2$ in equilibrium 
as a result of strangeness conservation. 
Using, on the other hand,  the standard parameterization for the 
BE-correlation function, one obtains
\be
\rho_2(q) = \rho_{11}(q)(1  + \lambda e^{-q^2 R^2})
\ee
and hence 
\be
F_2 = \frac{\int dq \rho_2(q)}{\int dq \rho_{11}(q)} = 1 + 
\lambda \int dq e^{-q^2 R^2} \rho_{11}(q) > 1,
\label{f2_false}
\ee
where the second term is only a very small correction of the order of a few
percent, given a typical source size of $R_{source} \simeq 5 \, \rm fm$
\cite{hbt_article}.  
Obviously the standard parameterization for the BE-correlation function is not
adequate for the case of rare particles subject to a conservation law. Since
the correlations due to strangeness conservation are not expected to introduce
any momentum dependence, we thus predict, that the BE-correlation 
for rare particles subject to $U(1)$ conservation should
asymptotically approach a value of 1/2, i.e. 
\be
C_{BE}(q \gg 1/R_{source}) \simeq \frac{1}{2}.
\ee 

Therefore, the rather difficult measurement of kaon correlations at 
heavy-ion collisions at 1 AGeV would not only provide information 
about the size of
the kaon emitting source but, more importantly, would also 
be a direct measurement
of the degree of equilibrium reached in these collisions.

\section{Conclusion}

 In this work, we addressed the question of number fluctuations of rarely
 produced particles.
 We carried out the analysis
 by solving the master equation derived in Ref.\cite{kkl}.
 The most important aspect of the master equation (\ref{generaln})
 is that we can treat the conservation laws that govern the rare particles
 exactly.  For instance, the strangeness conservation for kaon production at
 the SIS energy can be treated in this way.  Comparing the results of sections
 \ref{sec:with_u1} and \ref{sec:no_u1} certainly shows the difference
 such a constraint makes.
 In previous papers, some of us explored the consequence of requiring
 the exact conservation on the behavior of the average multiplicity
 in equilibrium as well as in evolving systems.
 In this work, we investigated the time evolution of the second factorial
 moment $F_2 = \ave{N(N-1)}/\ave{N}^2$ to explore
 the possibility of using it as
 a non-equilibrium measure.

 To cover a wide range of physical phenomena, we studied two extreme
 cases. (i) The initial population of the rare particle is much larger than
 the equilibrium population.  (ii) The initial population is much smaller.
 Our main conclusion is that the measurement of $F_2$ can certainly tell us
 if the equilibrium has not been reached.
 Moreover,
 the approach of the second
 factorial moment towards the equilibrium depends very much on the initial
 condition.  Assuming that the equilibrium population $\epsilon$
 is very small, we see that
 the smaller initial population results in the approach from above to the
 equilibrium value $(F_2^{\rm eq.} = 1/2 + {\cal O}(\epsilon))$. 
On the other hand
 a larger initial population results in the approach from below with a
 long period of very small $F_2$.
 Hence, the experimental value of $F_2$ can immediately tell us if the
 initial population was smaller ($F_2^{\rm exp} > 0.5$),
 larger ($F_2^{\rm exp} < 0.5$, possibly $F_2^{\rm exp} \ll 0.5$)
 or the system has already reached the
 equilibrium before the freeze-out ($F_2^{\rm exp} \simeq 0.5$).

In summary: the essential point of this work is  that in a system
in thermal equilibrium the average particle number fixes the
fluctuations. In case of rare particles subject to $U(1)$ charge conservation
these fluctuations are different from the simple Poisson type
finite number fluctuations and thus provide a measure of the
degree of equilibration reached in a system. This measurement can either be
achieved by measuring the inclusive one and two particle densities or via the
well known Bose-Einstein correlations.

\section*{Acknowledgment}
\hspace*{\parindent}
We thank Z. Lin for discussions.
This work was supported by the Director, Office of Energy Research,
Office of High Energy and Nuclear Physics, Division of Nuclear Physics,
the Office of Basic Energy
Science, Division of Nuclear Science, of the U.S. Department of Energy under
Contract No. DE-AC03-76SF00098,
the Committee for Scientific Research (KBN-2P03B 03018), and NSFC
under project 19928511.
S.J's work was supported in part by the Natural Sciences and Engineering
 Council of Canada and by le Fonds pour la Formation de Chercheurs et
 l'Aide \`a la Recherche du Qu\`ebec.

\appendix

\section{Initial time evolution for small initial particle numbers}
\label{app:init_time}

The leading time dependence of the  probabilities $P_n(\tau)$ can be
obtained by Taylor expansion around the initial time $\tau = 0$
\be
P_n(\tau) =
P_n(\tau=0) + \sum_{m}
\left.\frac{1}{m!}\frac{d^m P_n}{d\tau^m}\right|_{\tau=0}
\tau^m.
\ee
The time derivatives can be obtained by iteratively applying
Eq.\raf{generaln}.
To order $\tau_\alpha^3$ one obtains
\be
\ave{N(N-1)} &=&
\epsilon^2 \alpha^2 \left( (2 \tau_\alpha + \tau_\alpha^2) - \alpha (5 \tau_\alpha^2 -
\frac{5}{3}\tau_\alpha^3) + {\cal O}(\alpha^2) \right) \nonumber \\
\ave{N}^2 &=& \epsilon^2 \alpha^2 \left(
(1+\tau_\alpha)^2 - \alpha(2 \tau_\alpha + 3 \tau_\alpha^2 + \tau_\alpha^3)+ 
{\cal O}(\alpha^2) \right)
\non
\label{approx_1}
\ee
where we have neglected higher orders in the small variable
\be
\alpha &\equiv& \frac{N_0}{\epsilon} \simeq \frac{N_0}{\ave{N_{\rm eq.}}} 
\ll 1.
\\
\ee 
We have also rescaled the time $\tau_\alpha$ according to 
\be
\tau = \tau_\alpha \, \alpha. 
\ee 
Initially, at $\tau=0$, the
factorial moment starts out at zero 
\be F_2(\tau=0) =
\frac{\ave{N(N-1)}}{\ave{N}^2} = 0. 
\ee 
However, after a very short
time of the order of $\tau =\frac{N_0}{N_{\rm eq.}} $ corresponding
to $\tau_\alpha =1$ the factorial moment assumes a value \be
F_2(\tau_\alpha = 1) \simeq \frac{3}{4} \ee which is larger than
the final equilibrium result
 \be F_2(\tau\rightarrow \infty)
\simeq \frac{1}{2}. \ee 
Therefore, one expects that   the
factorial moment $F_2$ reaches a maximum value of about $F_2
\simeq 1$ after time of the order of $\tau = \frac{N_0}{N_{\rm
eq.}}$. Furthermore, $F_2$ approaches equilibrium from above thus
a measurement of $F_2
> 1/2$ will indicate that equilibrium has not been reached in a
heavy ion collision.

\section{Perturbative Solution}
\label{app:pert}

To solve (Eq.\raf{eq:gen_func})
\be
{\partial g \over \partial \tau} =
(1-x)\left( x g'' + g' - \epsilon g \right)
\label{eq:app_gen_func}
\ee
perturbatively, we first make an ansatz
\be
g(\tau, x) = g_{\rm eq.}(x) +
\sum_{n=1}^\infty \, e^{-n^2\tau} a_n \, h_n(x)
\;
\label{eq:app_gh}
\ee
with
\be
h_n = h_n^{(0)} + \epsilon h_n^{(1)} + \epsilon^2 h_n^{(2)} + \cdots
\label{eq:app_hn}
\ee
By substituting the expression \raf{eq:app_gh} into
Eq.\raf{eq:app_gen_func} and
collecting the same powers of $\epsilon$,
one can easily show that the equation for $h_n^{(0)}$ is
\be
(1-x)\left(
x h_n^{(0)\prime\prime} + h_n^{(0)\prime}
\right)
+n^2 h_n^{(0)} =  0
\label{eq:hn0eq}
\ee
which has the solution
 \be
 h_n^{(0)}(x)
 \equiv  F(n, -n; 1; x)
 =
 \sum_{s=0}^n {\prod_{i=0}^{s-1}(i^2 - n^2) \over (s!)^2} x^s
 \label{eq:hn0}
 \ee
 where $F(n, -n; 1; x)$ is a hypergeometric function.
 All other $h_n^{(s)}$'s are determined by the functional relations
\be
-n^2 h_n^{(s)} = (1-x)\left(
x h_n^{(s)\prime\prime} + h_n^{(s)\prime} - h_n^{(s-1)}
\right)
\ee

Keeping terms up to $\epsilon$ and $e^{-4 \tau}$
yields the following expressions for the relevant averages
\be
\ave{N}
& = &
g'(\tau, 1)
\non
& = &
\ave{N}_{\rm eq.} +
   {{\left( -1{+}{\frac{\epsilon}{5}} \right) a_1}{{e^{-\tau}}}}
   {+}
   {{\left( 2{-}{\frac{2\,\epsilon}{65}} \right) a_2}{{e^{-4\tau}}}}
   \non
\label{eq:pert_n}
   \\
\ave{N(N{-}1)} & = &
g''(\tau, 1)
\non
& = &
\ave{N(N{-}1)}_{\rm eq.} - {{{\frac{2\,\epsilon}{5}}  a_1}{{e^{-\tau}}}}
   {+}{{\left( 6{+}{\frac{2\,\epsilon}{13}} \right) a_2}{{e^{-4\tau}}}}
   \non
\label{eq:pert_nn_1}
\ee

To make a statement on how the averages behave as a function of
time, one needs to know the coefficients $a_n$. Obviously, the
initial condition fixes these coefficients \be g(0, x) = g_{\rm
eq.}(x) + \sum_{n=1}^\infty \, a_n \, h_n(x) = \sum_{s=0}^\infty \,
x^s P_s(0) \ee In general, to get $a_n$, one needs to invert \be
P_s(0)
 =
\sum_{n=s}^\infty A_{sn} \, a_n +{\cal O}(\epsilon) \label{eq:PnInAsn}
\ee where \be A_{sn} = {1\over (s!)^2} {\prod_{i=0}^{s-1}(i^2 -
n^2)} \label{eq:DefAsn} \ee It is not a trivial task in general to
solve the above equation for all $a_n$. A simple procedure to
solve for $a_n$'s can be given only if we ignore the ${\cal O}(\epsilon)$
corrections and also if there is a last index $N$ for which $P_N$
is non-zero. In that case, we can write \be P_s(0)
 =
\sum_{n=s}^N A_{sn} \, a_n
\label{eq:ZerothPs}
\ee
This is a triangular linear system of equations and can be easily solved by
first getting $a_N = P_N(0)/A_{NN}$ and then $a_{N-1}$ and so on.

To have a general understanding of how $a_n$'s behave, first
consider the condition \be 1 = \sum_{s=0}^\infty P_s(0) & = &
\sum_{s=0}^\infty \sum_{n=s}^\infty A_{sn} \, a_n \non & = &
\sum_{n=0}^\infty a_n \sum_{s=0}^n A_{sn} \ee From
Eqs.~(\ref{eq:hn0eq}), (\ref{eq:hn0}) and (\ref{eq:DefAsn}), it is
easy to see that \be \sum_{s=0}^n A_{sn} = h_n^{(0)}(1) =
\delta_{0n} \ee Hence, the above condition constrains \be a_0 = 1
\ee Otherwise, $a_n$'s only have to make $ 0 \le P_s(0)$, or \be 0
\le \sum_{n=s}^\infty A_{sn} \, a_n \ee If $s$ is odd then $A_{sn}
< 0$ for all $n \ge s$ and if $s$ is even then $A_{sn} > 0$ for
all $n \ge s$. To keep the probabilities positive, $a_n$'s must
have alternating signs starting with $a_1 < 0$ and $\absol{a_n}$
must be a monotonic decreasing function of $n$.  Furthermore, to
keep the probabilities finite, $\absol{a_n}$ must decrease faster
than any power of $n$.

Numerical investigation shows
that the size of $a_s$ is ${\cal O}(1)$ up to $s \simeq \sqrt{N_0}$
and from then on $\absol{a_s}$ falls like a Gaussian (faster for larger
$N_0 \gsim 15$).
Empirically,
\be
a_n = (-1)^n 2 \exp\left(-n^2/M_0\right) \ \ \ \ (n \ge 1)
\label{eq:emp_an}
\ee
where $M_0 \approx N_0$ works up to 
$N_0 \simeq 15$.  
For larger $N_0$, some small $P_s$
can become negative.
With $M_0 = 15$, the above expression (\ref{eq:emp_an})
gives
\be
& &
a_1 = -1.87101\ \ \  a_2 = 1.53186\ \ \ a_3 = -1.09762
\non
& &
a_4 = 0.688308\ \ \ a_5 = -0.377751
\ee
These coefficients result in a probability distribution with
$\ave{N}_{\tau =0}=14.8$ and 
$\ave{N(N{-}1)}_{\tau = 0} =  210.2$ as shown in
Fig.~\ref{fig:EmpPn}.

Using the initial distribution given by a Poisson distribution
with $N_0 = 15$ and solving Eq.~(\ref{eq:ZerothPs}) yield
\be
& &
a_1 = -1.86667\ \ \ a_2 = 1.52\ \ \  a_3 = -1.08444,
\non
& &
a_4 = 0.682074\ \ \ a_5 = -0.380919
\ee
These of course result in $\ave{N(N{-}1)}_{\tau=0} =  N_0^2 = 225$.
\begin{figure}[htb]
 \setlength{\unitlength}{1cm}
  \epsfxsize=8cm
  \centerline{\epsfbox{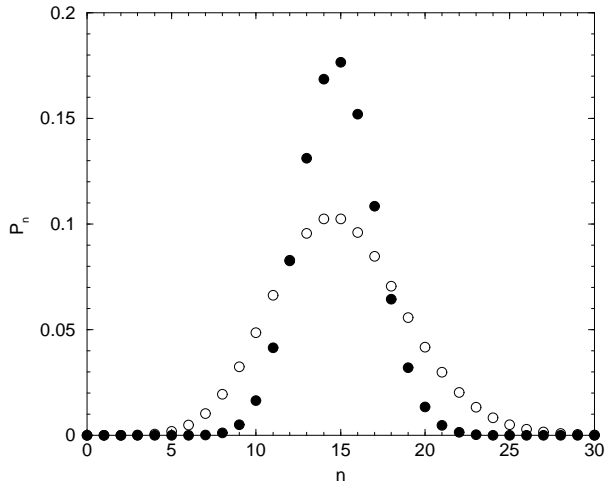}}
  \caption{The filled circles represents the probability distribution
  resulting from using
  Eq.~(\protect\ref{eq:emp_an}) with $M_0 = 15$.
  The open circles represents a Poisson distribution with
  $N_0 = 15$.
  }
  \label{fig:EmpPn}
\end{figure}

For completeness we also quote the results with a Poissonian initial
distribution with 
$N_0= 5$,
solving Eq.~(\ref{eq:ZerothPs}) yields
\be
& &
a_1 = -1.6027\ \ \  a_2 = 0.871375\ \ \  a_3 = -0.347491
\non
& &
a_4 = 0.107955\ \ \  a_5 = -0.0272918
\ee
These values are used to calculate the perturbative solutions shown in
Fig.~\ref{fig:pert_sol}.
The Gaussian formula gives
\be
& &
a_1 = -1.63746\ \ \  a_2 = 0.898658\ \ \  a_3 = -0.330598
\non
& &
a_4 = 0.0815244\ \ \  a_5 = -0.0134759
\ee


{}

\end{document}